\documentstyle[pra,aps,twocolumn]{revtex}

\title{Comment on: necessary and sufficient condition of separability of any system}
 \author{T. Eggeling, K.G.H. Vollbrecht, M.M. Wolf
  \\[1ex]
  {\small Institut f{\"u}r Mathematische Physik, TU Braunschweig,}\\
  {\small Mendelssohnstr. 3, 38106 Braunschweig, Germany.}}
\date{\today}

\newtheorem{The}{Theorem}
\begin{document}
\draft \maketitle\begin{abstract} In a recent paper
(quant-ph/0102133) Chen, Liang, Li and Huang suggest a necessary
and sufficient separability criterion, which is supposedly
practical in judging the separability of any mixed state. In this
note we briefly recapitulate their main result and show that it is
a reformulation of the problem rather than a practical criterion.

\end{abstract}

\narrowtext\subsection{The content of the paper}

Deciding whether a given quantum state is entangled or not is a
central problem in quantum information theory. For the smallest
non-trivial systems (with $2\times 2$ resp. $2\times 3$
dimensional Hilbert spaces) the positivity of the partial
transpose is known to be an efficient necessary and sufficient
criterion \cite{hor}. Beyond these special cases however, no such
calculable criterion is known. For this reason a paper titled {\it
necessary and sufficient condition of separability of any
system}\cite{sep} should be of interest for the whole quantum
information community (this is one reason, why we post this
comment on the quant-ph server).

For the benefit of other readers of the archive, we recapitulate
the main result in \cite{sep} and show that it is nothing but a
reformulation of the definition of separability --- which is
naturally a necessary and sufficient criterion for itself.

Before we give a slightly reformulated but textual unchanged version of the main result
in \cite{sep} let us introduce the following two-dimensional projectors:
\begin{equation}\label{Pij}
P_{ij}=|i\rangle\langle i| + |j\rangle\langle j| \quad i\neq j.
\end{equation}
Such a projection can be understood, as a projection from a larger
Hilbert space ${\bf C}^n$ to the qubit space ${\bf C}^2$. Although
the results in \cite{sep} are not expressed in terms of these
projectors, we will make use of them in order to avoid lengthy and
cumbersome notations. Any proposition in \cite{sep} has then a
simple and rather intuitive translation. For example, take a
vector $|\Psi\rangle =\sum_{ij }  A_{ij} |ij\rangle$. To say that
$\{A_{11},A_{12}\}$ is parallel to $\{A_{21},A_{22}\}$ is now
 equivalent to the fact that the respective two qubit projection $P_{12}\otimes P_{12}|\Psi\rangle$
is separable \cite{B}. The main theorem in \cite{sep} then reads
as follows:

 \begin{The}Let $\rho$ be a mixed state on ${\bf C}^n\otimes{\bf
C}^m$ with eigenvectors $\{x_k\}$ (unnormalized). Let $$\rho^
{ijkl}=P_{ki}\otimes P_{lj} \rho P_{ki}\otimes P_{lj} $$ denote
the projection of $\rho$ to a two qubit space. Then $\rho$ is
separable iff the following two conditions hold:

\begin{enumerate}
\item ´For all two-qubit projections $ \rho^{ijkl  }$ the concurrence
(denoted as $a^r$) as introduced by Wootters \cite{Wootters} is
negative (or zero, if we take the original definition
$c=\max\{0,a^r\}$) \cite{cond1}.

\item There exists an isometry $U$ corresponding to a
decomposition $\rho= \sum _k |z_k\rangle \langle z_k|$ with
$|z_k\rangle=\sum_j U_{kj}|x_j\rangle$, such that every two qubit
projection $P_{ki} \otimes P_{lj} |z_k\rangle$ is a product vector
\cite{cond1,cond2}.

\end{enumerate}
\end{The}

\subsection{What it is about}

Now let us have a closer look at the two conditions in Theorem 1.
The first condition utilizes Wootters formula for the concurrence
as separability criterion for the mixed two qubit states (even the main part of Wootters proof is repeated).
However, this is rather confusing since their final result is
completely independent of the choice of the separability criterion
at this point. One could equivalently take the partial
transposition criterion.

 Condition 2 is equivalent to saying that
all the vectors $|z_k\rangle$ have to be separable (see page 2 in \cite{sep}). But then the
search for the transformation $U$ is nothing but a search for a
decomposition into product vectors, i.e., just a rephrasing of the
initial problem. Notwithstanding the fact that condition 1 is
therefore redundant it is not even a good necessary criterion for
separability, since the usual PPT-criterion is both, easier to
calculate and stronger.

The necessary and sufficient separability criterion, which is
``{\it practical in judging}'' thus turns out to be a tautology.
The fact that it works for an example of a bound entangled state
is not surprising, since all the known examples of such states are
just constructed in a way that it is easily seen that they are
entangled (i.e., they do not have any product vector in their
range).


\begin{references}
\bibitem{B} For other readers of the paper it might also be useful
to note that the matrices $B^r$ are just $\sigma_y\otimes\sigma_y$
on the respective two qubit space.
\bibitem{cond1} Originally, in \cite{sep} only projections of the
form $\rho^{ij11}$ were considered. This would make Theorem 1
false, since both conditions would then be satisfied for {\it any}
$\rho$ such that $\rho({\bf 1} \otimes |1\rangle \langle1|)=0$.
Therefore we extended the set of projections. This is however not
the main point of our criticism.
\bibitem{cond2} The formulation of this condition in \cite{sep} is
slightly different. They require that the intersection of certain
sets of isometries is not empty. However, any element of this
intersection is just characterized by the properties of $U$.
\end{references}
\end{document}